\documentclass{article}
\usepackage[utf8]{inputenc}
\usepackage{amssymb}
\usepackage{wrapfig}
\usepackage{framed}
\usepackage{graphicx}
\usepackage{subfigure}
\usepackage{tabularx}
\usepackage[hyphens]{url}
\usepackage{hyperref}
\usepackage{color}
\usepackage{amsmath}
\usepackage{booktabs}
\usepackage{tikz}
\usepackage{tikz-qtree}
\usepackage{mathptmx}
\usepackage{algorithm2e}
\newtheorem{theorem}{Theorem}
\newtheorem{proof}{Proof}

\newcommand{\is}[1]{\makebox[0pt]{#1}}
\newcommand{\card}[1]{\left|#1\right|}

\makeatletter
\newcommand*{\defeq}{\mathrel{\rlap{%
  \raisebox{0.3ex}{$\m@th\cdot$}}%
  \raisebox{-0.3ex}{$\m@th\cdot$}}%
  =}
\makeatother

\newcommand{\calE}{\mathcal{E}}

\newcommand{\calP}{\mathcal{P}}

\newcommand{\calW}{\mathcal{W}}

\newcommand{\id}{\mathrm{id}}
\newcommand{\name}{\mathrm{name}}
\newcommand{\gtin}{\mathrm{GTIN}}
\newcommand{\cas}{\mathrm{CAS}}
\newcommand{\doi}{\mathrm{DOI}}
\newcommand{\myemail}{\mathrm{email}}

\newcommand{\fr}{\mathit{freq}}

\makeatletter
\def\@copyrightspace{}
\makeatother

\usepackage{enumitem}

\newcommand{\html}[1]{\textsc{#1}}
\renewcommand{\paragraph}[1]{\noindent\textbf{#1.}}

\newcommand{\ignore}[1]{}

\begin{document}
\title{Harvesting Entities from the Web\\ Using Unique Identifiers -- IBEX\\[0.5cm]
Extraction des entités du Web\\à l'aide d'identifiants uniques -- IBEX\\[0.5cm]}
\author{
Aliaksandr Talaika\\
Max Planck Institute for Informatics, Germany\\[0.4cm]
Joanna Biega\\
Max Planck Institute for Informatics, Germany\\[0.4cm]
Antoine Amarilli\\
T\'el\'ecom ParisTech, France\\[0.4cm]
Fabian M. Suchanek\\
T\'el\'ecom ParisTech, France
}
\date{May 2015}
\maketitle
\newpage
\begin{abstract}
In this paper we study the prevalence of unique entity identifiers on the Web.
These are, e.g., ISBNs (for books), GTINs (for commercial products), DOIs (for
documents), email addresses, and others. We show how these identifiers can be
harvested systematically from Web pages, and how they can be associated with
human-readable names for the entities at large scale.

Starting with a simple extraction of identifiers and names from Web pages, we
show how we can use the properties of unique identifiers
to filter out noise and clean up the extraction result on the entire corpus.
The end result is a database of millions of uniquely identified entities of different types, with an accuracy of 73--96\% and a very high coverage compared to existing knowledge bases.
We use this database to compute novel statistics on the presence of products, people, and other entities on the Web.\\

\noindent This work was published at WebDB 2015 \cite{ibex}.\\

\begin{center}
\textbf{Résumé}\\
\end{center}
Ce travail s'intéresse aux identifiants uniques sur le Web. Ceux-ci incluent,
entre autres, les ISBN (pour les livres), les GTIN (pour des produits
commerciaux), les DOI (pour des documents), et les adresses de courriel. Nous
montrons une méthode systématique pour extraire ces identifiants des
pages Web, et pour les associer à des noms lisibles.

Nous présentons d'abord une extraction simple des identifiants et des noms, et
nous montrons ensuite comment les propriétés spécifiques des identifiants
permettent d'éliminer le bruit et de nettoyer le résultat de l'extraction sur la
totalité du corpus. Le résultat final est une base de plusieurs millions
d'entités uniques, avec une précision de 73--96\%, et une couverture très large
comparée à celle d'autres bases de connaissances. Nous utilisons cette base de
données pour calculer de nouvelles statistiques sur les présence des gens, des
produits, et d'autres entités sur le Web.\\

\noindent Ce travail a été publié à WebDB 2015 \cite{ibex}.

\end{abstract}
\newpage
\section{Introduction}

\begin{figure}[h!]
\centering

\frame{\includegraphics[width=\columnwidth]{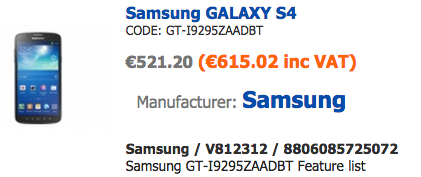}}
\caption{A Web page snippet about a product}\label{example}
\end{figure}

\paragraph{Unique ids} The Web is an almost endless resource of named entities, such as commercial products, people, books, and organizations. In this paper, we focus on those entities that have unique \emph{ids}. An id is any string or number that distinguishes the entity in a globally unique way from other entities.
For example, commercial products have ids in the form of GTINs. These are the numeric codes printed below the bar code on the package or item. They also frequently appear on the Web. Figure~\ref{example} shows an excerpt from a Web page about a commercial product. The GTIN (8806085725072) appears at the bottom right.

But not just commercial products have ids. A surprisingly large portion of other entities also do. Companies have tax identification numbers; books have ISBNs; documents have document identifiers; chemical substances have ids in the form of CAS registry numbers, and so on. Quite frequently, Web pages that talk about these entities also mention their ids.

\paragraph{Goal} Our goal is to harvest these ids at large scale from the Web, together with the names of the entities that they identify. That is, our goal is to build a database that contains, in the example, $\langle$8806085725072, Samsung Galaxy S4$\rangle$. Using named entity recognition (NER), ids and entity names can be spotted in the pages. However, a page usually contains several entity names, and only one of them is usually the name of the entity in question. The challenge is thus to associate, with each id, the proper name for the entity. In the example, the challenge is to find that the correct name for the id ``8806085725072'' is ``Samsung Galaxy S4'' -- and not ``Samsung'', ``VAT'', or ``GT-I9295ZAADBT''.

It is far from trivial to associate the correct entity name to an id. First, Web pages contain usually dozens of entity names, so it is not clear  which one corresponds to the id. In the example, ``Samsung'' is clearly an entity name, but not the correct one. Worse, some Web pages contain several ids and several entity names at the same time, so we must correctly match the ids and names on the page. The excerpt of Figure~\ref{example} is taken from a page that lists dozens of Samsung products.

Finally, if we want to find entity ids and names at Web scale, we need an approach that is both fast and resilient. It must run on hundreds of millions of Web pages, and it must accept entirely arbitrary pages, with possibly erroneous content, broken structure, or noisy information.
This makes it impossible to rely on wrapper induction, or indeed on any predefined or learnable DOM tree structure. We have to be able to find the entity names in tables, in lists, as well as in plain unstructured text. These challenges come in addition to the usual difficulties such as non-standard HTML code, non-semantic markup (e.g., tables used for page layout), and creative tag combinations to arrange tabular information.

\paragraph{Contribution} In this paper, we show how to systematically collect unique ids from Web pages, and how to associate each id to the correct entity name.
We first use vanilla NER methods to extract ids and candidate names from each Web page. Then, we rely on the inherent characteristics of unique identifiers to filter the name candidates so as to keep only the correct names for the entities. Our method is scalable, fast, and resilient enough to run on arbitrary Web pages.

This allows us to extract millions of distinct entities from the Web, with an accuracy of 73\% to 96\% depending on the entity types. The result is a database of entity ids and names, with information about which pages mention which entities. The crucial advantage of our database is that every entity is guaranteed to be \emph{unique},
so we can count \emph{distinct} entities without being biased by duplicates. Thus, we can perform a detailed study of entities that exist on the Web: we can identify Web sites that are hubs for books or documents, we can build statistics about frequent first names of people, and we can determine which countries produce most products. We can trace producing countries, importing countries, and the flow of products from one to the other. In other words, we show not only how entities with unique ids can be extracted from the Web, but also how they are distributed on the Web, and in the world.

Our contributions are:
\begin{itemize}[noitemsep,nolistsep,leftmargin=0.3cm,midpenalty=0,label=$\bullet$]
\item The paradigm of \emph{id-based entity extraction} (IBEX), harvesting entity ids with their names at large scale from the Web.
\item A database of unique entities with millions of objects
  and an accuracy of 73--96\%.
\item Detailed analyses about the distributions of these objects.
\end{itemize}

\noindent The paper is structured as follows. We first discuss related work in
Section~\ref{sec:related-work}. We then define our problem in
Section~\ref{sec:problem}, and our approach to solve it in
Section~\ref{sec:approach}. The details of Web page parsing are described in Section~\ref{sec:parsepage}.
We present our experiments in Section~\ref{sec:experiments}, and show in
Section~\ref{sec:analyses} some examples of analyses that can be conducted on
our entity database. Section~\ref{sec:conclusion} concludes. This work was published at WebDB~2015~\cite{ibex}.
\section{Related Work}
\label{sec:related-work}

Our goal is to extract unique ids from Web pages and to associate, with each id, the correct entity name.
We now survey related work about this goal.

\subsection{Information Extraction (IE)}

\paragraph{Named Entity Recognition}
Named entity extraction (NER) is the task of recognizing and categorizing real-world entities in textual resources (see \cite{sunita,1683775.Web.IE.Survey} for surveys). We rely on NER to spot ids and candidate names in Web pages. However, our focus is not on these methods. Rather, our focus is on associating, with each id, the correct name among several extracted name candidates. While NER can spot entity names, it cannot determine which entity name relates to which id, if the page contains several entity names. Our approach aims at solving this problem.

\paragraph{Wrapper induction} Wrapper induction \cite{42,61,65,10,18,nilesh2011,nora,mirko,gulhane} learns the structure of a Web page and produces a so-called \emph{wrapper}, which can then be applied to extract information from other Web pages of the same form. These approaches exploit the fact that large Web sites are typically generated from a source by the help of templates.
In our setting, we cannot make such an assumption, because we target arbitrary Web pages of arbitrary sites. We may have only a handful of pages for each site, plus a large number of pages that do not belong to any large site. The DIADEM project~\cite{diadem} can deal with such variety, but targets the deep Web, while we target the surface Web.

\paragraph{Structured IE} A large suite of approaches (e.g., \cite{59,20,25,26}) aims at extracting information from structured sources. Some techniques use visual clues \cite{40,60}; others make use of the DOM structure of the Web pages \cite{41}. Yet others make use of the schema \cite{2}.
Unlike these approaches, our method does not assume any particular structure in Web pages. It does not require that pages resemble each other, it does not need training data, and it does not assume a given schema. Rather, it works on both structured and completely unstructured sources across arbitrary sites and domains.

\paragraph{Product extraction}
One of the applications of our work is the extraction of commercial products. Previous work on product extraction focused on matching product offers to products and their attributes \cite{Kannan:2011:MUP:2020408.2020474,Kopcke:2012:TER:2247596.2247662}. The work by \cite{Kopcke:2012:TER:2247596.2247662} succinctly mentions also manufacturer extraction from product titles. Supervised learning approaches have been proposed to update product catalogues using new offers \cite{Nguyen:2011:SPO:1988776.1988777}, or to determine product prices \cite{Agrawal:2012:AWO:2339530.2339602}. These approaches, however, build on an existing catalog of products.
This is because these KBs focus more on popular entities than on the long tail.
Our goal, in contrast, is the creation of such a catalog.

Other work \cite{Putthividhya:2011:BNE:2145432.2145598,Probst:2007:SLA:1625275.1625732,Ghani:2006:TMP:1147234.1147241} handles product attribute extraction. \cite{Pham:2011:SSL:2069216.2069254} gives a method to discover product information regions on Web pages. While these approaches share our goal of
extracting product data from Web pages, they
do not target the creation of a global database of unique products, as we do.

\paragraph{Knowledge bases} Recent work has led to the automated creation of large knowledge bases \cite{yago,dbpedia,nell,textrunner}. These contain many popular and important entities, but do not aim to be exhaustive. DBpedia, e.g., contains ``only'' a few ten thousand instances of the class \emph{Product}, most of them being named ships. Our goal, in contrast,
is to collect products systematically from the Web.

\paragraph{Web-scale databases} Several projects \cite{webtables,96,19,entitycube,statsnowball} construct a queryable database of Web objects. While we share this goal, we use ids to achieve it.
This has two advantages.
First, we can extract even from Web pages with poor structure, or no structure at all.
Second, we build a database of \emph{unique} entities, where each entity is guaranteed to appear at most once.

\subsection{Entity Databases} \label{codedb}

There are several databases of unique Web entities.

\paragraph{Products} The UPC Database (\url{http://www.upcdatabase.com}) contains 1.6M ids of commercial products, but is not available for download. GTIN13.com (\url{http://gtin13.com}) and Smoopa (\url{http://www.smoopa.com/}) are other databases of ids, but no information on their content is freely available. Amazon and other booksellers store large numbers of ISBN codes, and some search engines may rely on a catalog of products, but those databases are not available for download, because having a product repository is a competitive advantage.
Our open methodology and dataset, in contrast, can be used freely by any vendor to improve coverage.

\paragraph{Documents} The International DOI Foundation (\url{http://www.doi.org/}) assigns identifiers to text documents upon request. There are 84M document ids. However, the foundation does not provide a central search capability across all DOI names, and the data cannot be downloaded.

\paragraph{Chemistry} The Chemical Abstracts Service (CAS) (\url{http://www.cas.org}) maintains a registry of more than 71M organic and inorganic substances. However, this data is not available for free. The Common Chemistry Website  (\url{http://www.commonchemistry.org/}) provides publicly available data, but for only 7,900 substances.

\paragraph{People} There are some commercial Websites that scrape the Web for personal data (e.g. \url{http://www.yasni.com/}); however, they do not provide a downloadable dataset of people. Social networks, likewise, harvest personal data, but do not
make them available for public use.

The main advantage of our database that it is freely available, while most existing databases are commercial. Furthermore, our method is a general technique that can apply to any entities that have unique identifiers, whereas existing approaches are domain-specific.

\section{Problem Statement}
\label{sec:problem}

Our goal is to build a database of unique entity ids from Web pages, and to associate to each id a human-readable name.
We now formally define our notion of ids and the problem that we study.

\paragraph{Ids}
For us, an \emph{entity} is any real-world object such as a person, a book, a product model, or a shipping container.
An \emph{id} is a string that is used as an identifier for an entity.

There are different \emph{types} of ids, i.e., groups of ids that follow the same syntax, and that refer to entities of the same domain. For example, ISBNs are always sequences of 10-13 digits, and they are used to identify books. CAS numbers are sequences of 8 digits that identify chemical substances.
The only assumptions that we make is that no two entities can have the same id (e.g., one ISBN cannot refer to two different books), and that one entity can only have one id in one type (e.g., a book can only have one ISBN, but it can have also a GTIN, because GTINs are a different id type that happen to include books).

In some cases, we cannot make the assumption that entities have only one id in a type.
For example, every personal email address belongs to one person, but one person can have several personal email addresses.
In this case, we call the identifier a \emph{pseudo-id}.
Our approach can also collect entities by their pseudo-ids, but it cannot
guarantee the uniqueness of the collected entities in this case: for instance,
the same person may appear multiple times in the constructed database, under their
various email addresses.

We assume that every entity has one or several \emph{names}. A name is a human-readable string that identifies the entity intuitively.
\begin{table}[t]
\caption{Id types}
\label{otherids}
\centering
\begin{tabular}{ll}
\toprule
{\bf Id type} & {\bf Entities}\\
\midrule
ISBN  & Books \\
GTIN  &Products\\
CAS  &Chemicals\\
DOI  &Documents\\
VATIN  &Companies \\
BIC  &Banks \\
NSN  &Military products \\
ISIN  &Stocks \\
VIN  &Vehicles\\
GRID  &Digital recordings\\
ISAN  &Audiovisual material\\
Pub\#    &US Patents \\
ILU   &Containers\\
MESH  &Chemicals\\
OMIM  &Diseases\\
ICD-10  &Diseases\\
Email & People/organizations (pseudo-id)\\
IBAN & People/organizations (pseudo-id)\\
Phone & People/organizations (pseudo-id)\\
\bottomrule
\end{tabular}
\end{table}

\begin{table*}[t]
\caption{Examples for ids and entity names}
\label{idexamples}
\begin{tabularx}{\linewidth}{llX}
\toprule
{\bf Id type} & {\bf Id} & {\bf Entity name}\\
\midrule
GTIN & 00068888883955	& Pyramid PA305 100w Rack Mount Amplifier with Mixer\\
GTIN & 09783540442820 & 	Machine Translation: From Research to Real Users\\
CAS&	78123-16-7&	N-benzyl-2-(2-methyl-1H-indol-3-yl)acetohydrazide\\
CAS&	67011-42-1&	3-acetamido-5-(hexanoylamino)-2,4,6-triiodo-benzoic acid\\
DOI & 10.1037/a0024143	& Cognitive niches: An ecological model of strategy selection.\\
DOI&	10.2136/sssaj198... &	A Simple Method for the Estimation of Calcium and Magnesium Carbonates\\
Email & widom@cs.stanford.edu & Dr. Jennifer Widom\\
\bottomrule
\end{tabularx}
\end{table*}

\paragraph{Examples} The notion of ids and id types is a very general one, so
our approach will apply to a large variety of domains.
Table~\ref{otherids} presents examples of id types. They cover entities that are
intrinsically Web-based, such as Web documents, but also a large number of
real-world entities, such as chemicals, commercial products, vehicles, books,
magazines, and many more. Our experiments in this paper will focus on the following id types:
\begin{itemize}[noitemsep,nolistsep,leftmargin=0.4cm,midpenalty=0,label=$\bullet$]
\item \emph{Global trade item numbers} (GTINs) are identifiers for commercial products. They are the generalization of previous product ids such as UPCs, UCCs, and EANs. They also generalize ISBNs (for books). GTINs are assigned by the GS1, an international standards body, and can identify anything from digital cameras and kitchen appliances to books, toys, and pencils.
\item \emph{CAS numbers} are identifiers for chemical substances. They are assigned by the Chemical Abstracts Service to all substances described in the open scientific literature.
\item \emph{Digital object identifiers} (DOIs) are used to identify electronic documents such as PDF files. A DOI takes the form ``prefix/suffix''. The DOI Foundation assigns the prefixes centrally to registered document providers, and
the providers assign the suffix locally to the documents they produce.
\item \emph{email addresses} as pseudo-ids for people.
\end{itemize}

\noindent All of these id types have in common that
there is no structured open registry of all their entities (see
Section~\ref{codedb}).
Table~\ref{idexamples} gives real examples of entities of
these types, obtained from our data.

\paragraph{Problem statement}
The input to our method is a set of \emph{Web pages} obtained from a Web
crawl. In addition, we are given an id type~$t$ and
two NER modules for~$t$: the \emph{id validator} $f_t^\id$  and the \emph{name finder} $f_t^\name$. The id validator is a function that takes as input a string and returns \emph{true} iff the string is an id of type~$t$.  The name finder is a function that, given an id of type~$t$ and a string, extracts possible candidate names for that id from the string.
These can be single tokens or multi-words.

The problem is that we will sometimes find multiple ids accepted by $f_t^\id$ on the same page, and $f_t^\name$ will typically extract a large number of name candidates. Our goal is to figure out which candidate belongs to which id, and what is the best name for which id. This is particularly difficult if a page talks about several entities, and thus contains several ids and several names. Even if the page just talks about one entity, it typically contains dozens if not hundreds of name candidates.
As we will show, the correct name can be selected with high precision by leveraging the uniqueness property of ids at large scale.

The output of our method is an \emph{entity database} for type $t$, which contains ids of type~$t$, and associates each id with the name of the entity -- much like Table \ref{idexamples}. In addition, we store with each entity the URLs of the Web pages where the entity was found.

\section{Approach}
\label{sec:approach}

We first describe our approach, and next discuss the coverage that we can expect from it.

\subsection{Method Description}

In all of the following, we assume given a set $\calP$ of Web pages, an id type~$t$, and NER modules $f_t^\id$ and $f_t^\name$ for that type.  Our approach proceeds in 3 phases.
We first describe our algorithm at a high level, deferring the implementation
of Phase~1 to Section~\ref{sec:parsepage} and more details to the appendix.

\paragraph{Phase~1} The first phase extracts ids and name candidates. Let us consider one page $p \in \mathcal{P}$. We first split the page into \emph{records} $r_1, \ldots, r_n$, where each record is a region of~$p$ that contains exactly one id. For each record $r_i$, we extract the id $id_i$ and all name candidates $name_{i,j}$, $j=1,...,m_i$.
To account for differences in writing, we normalize all name candidates by upper-casing them, and by retaining only letters (alphanumeric characters for CAS).
Each name candidate also comes with a score that indicates its likelihood of
being the correct name for $id_i$. We discuss different scoring models and our
final choice in Appendix \ref{scoremodels}.

The result of this process is a table of the following form for each record $r_i$:
\[R1_i^p \defeq \{\langle id_i, name_{i,j}, score_{i,j}, url(p) \rangle \mid j=1,...,m_i\}\]
Its rows contain the id, a name candidate for this id, a score for this name candidate, and the URL of the page.
Note that the same name may occur multiple times in $R1_i^p$ (with the same score or with different scores) if the same name was extracted multiple times in~$r_i$.
The output of the first phase is then the union of these tables:
\[R1 \defeq  \bigcup_{p \in \calP, r_i \in p} R1_i^p\]
Again, $R1$ will contain several name candidates for the same id,
extracted from the same page or from different pages. It may also contain the same name multiple times. The idea is that the subsequent phases will filter out the erroneous names.

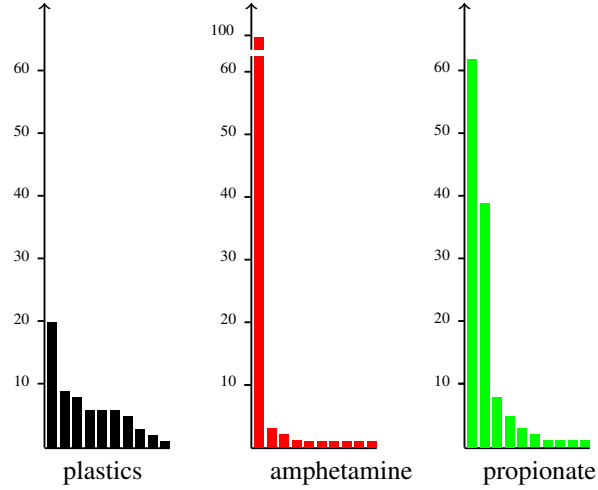
\begin{figure}
  \begin{center}


\begin{tikzpicture}
  \node[right,color=black, font=] at (0.675000, -6.188889) {plastics};
  \node[right,color=black, font=] at (3.425000, -6.188889) {amphetamine};
  \node[right,color=black, font=] at (6.258333, -6.188889) {propionate};
  \draw[->, draw=black, line width=0.7086614168764338] (0.550000, -5.850000) -- (0.550000, 0.066667);
  \draw[->, draw=black, line width=0.7086614168764338] (3.300000, -0.558333) -- (3.300000, 0.066667);
  \draw[->, draw=black, line width=0.7086614168764338] (6.133333, -5.850000) -- (6.133333, 0.066667);
  \draw[draw=black, line width=0.7086614168764338] (0.550000, -5.858333) -- (2.216667, -5.858333);
  \draw[draw=black, line width=0.7086614168764338] (3.300000, -5.858333) -- (4.966667, -5.858333);
  \draw[draw=black, line width=0.7086614168764338] (6.133333, -5.858333) -- (7.800000, -5.858333);
  \node[right,color=black, font=\fontsize{5.905512}{5.905512}\selectfont] at (0.016667, -4.963889) {10};
  \node[right,color=black, font=\fontsize{5.905512}{5.905512}\selectfont] at (0.016667, -4.130556) {20};
  \node[right,color=black, font=\fontsize{5.905512}{5.905512}\selectfont] at (0.016667, -3.297222) {30};
  \node[right,color=black, font=\fontsize{5.905512}{5.905512}\selectfont] at (0.016667, -2.463889) {40};
  \node[right,color=black, font=\fontsize{5.905512}{5.905512}\selectfont] at (0.016667, -1.630556) {50};
  \node[right,color=black, font=\fontsize{5.905512}{5.905512}\selectfont] at (0.016667, -0.797222) {60};
  \node[right,color=black, font=\fontsize{5.905512}{5.905512}\selectfont] at (2.766667, -0.797222) {60};
  \node[right,color=black, font=\fontsize{5.905512}{5.905512}\selectfont] at (2.766667, -1.630556) {50};
  \node[right,color=black, font=\fontsize{5.905512}{5.905512}\selectfont] at (2.766667, -2.463889) {40};
  \node[right,color=black, font=\fontsize{5.905512}{5.905512}\selectfont] at (2.766667, -3.297222) {30};
  \node[right,color=black, font=\fontsize{5.905512}{5.905512}\selectfont] at (2.766667, -4.130556) {20};
  \node[right,color=black, font=\fontsize{5.905512}{5.905512}\selectfont] at (2.766667, -4.963889) {10};
  \node[right,color=black, font=\fontsize{5.905512}{5.905512}\selectfont] at (5.608333, -0.788889) {60};
  \node[right,color=black, font=\fontsize{5.905512}{5.905512}\selectfont] at (5.608333, -1.622222) {50};
  \node[right,color=black, font=\fontsize{5.905512}{5.905512}\selectfont] at (5.608333, -2.455556) {40};
  \node[right,color=black, font=\fontsize{5.905512}{5.905512}\selectfont] at (5.608333, -3.288889) {30};
  \node[right,color=black, font=\fontsize{5.905512}{5.905512}\selectfont] at (5.608333, -4.122222) {20};
  \node[right,color=black, font=\fontsize{5.905512}{5.905512}\selectfont] at (5.608333, -4.955556) {10};
  \path[fill=black, line width=0.7086614168764338] (0.591667, -4.183333) rectangle (0.716667, -5.850000);
  \path[fill=black, line width=0.7086614168764338] (0.758333, -5.100000) rectangle (0.883333, -5.850000);
  \path[fill=black, line width=0.7086614168764338] (0.925000, -5.183333) rectangle (1.050000, -5.850000);
  \path[fill=black, line width=0.7086614168764338] (1.091667, -5.350000) rectangle (1.216667, -5.850000);
  \path[fill=black, line width=0.7086614168764338] (1.258333, -5.350000) rectangle (1.383333, -5.850000);
  \path[fill=black, line width=0.7086614168764338] (1.425000, -5.350000) rectangle (1.550000, -5.850000);
  \path[fill=black, line width=0.7086614168764338] (1.591667, -5.433333) rectangle (1.716667, -5.850000);
  \path[fill=black, line width=0.7086614168764338] (1.758333, -5.600000) rectangle (1.883333, -5.850000);
  \path[fill=black, line width=0.7086614168764338] (1.925000, -5.683333) rectangle (2.050000, -5.850000);
  \path[fill=black, line width=0.7086614168764338] (2.091667, -5.766667) rectangle (2.216667, -5.850000);
  \path[fill=red, line width=0.7086614168764338] (3.341667, -0.641667) rectangle (3.466667, -5.850000);
  \path[fill=red, line width=0.7086614168764338] (3.341667, -0.391667) rectangle (3.466667, -0.558333);
  \node[right,color=black, font=\fontsize{5.905512}{5.905512}\selectfont] at (2.633333, -0.297222) {100};
  \path[fill=red, line width=0.7086614168764338] (3.675000, -5.675000) rectangle (3.800000, -5.841667);
  \path[fill=red, line width=0.7086614168764338] (3.508333, -5.591667) rectangle (3.633333, -5.841667);
  \path[fill=red, line width=0.7086614168764338] (3.841667, -5.758333) rectangle (3.966667, -5.841667);
  \path[fill=red, line width=0.7086614168764338] (4.008333, -5.766667) rectangle (4.133333, -5.850000);
  \path[fill=red, line width=0.7086614168764338] (4.175000, -5.766667) rectangle (4.300000, -5.850000);
  \path[fill=red, line width=0.7086614168764338] (4.341667, -5.766667) rectangle (4.466667, -5.850000);
  \path[fill=red, line width=0.7086614168764338] (4.508333, -5.766667) rectangle (4.633333, -5.850000);
  \path[fill=red, line width=0.7086614168764338] (4.508333, -5.766667) rectangle (4.633333, -5.850000);
  \path[fill=red, line width=0.7086614168764338] (4.675000, -5.766667) rectangle (4.800000, -5.850000);
  \path[fill=red, line width=0.7086614168764338] (4.841667, -5.766667) rectangle (4.966667, -5.850000);
  \path[fill=green, line width=0.7086614168764338] (6.175000, -0.683333) rectangle (6.300000, -5.850000);
  \path[fill=green, line width=0.7086614168764338] (6.341667, -2.600000) rectangle (6.466667, -5.850000);
  \path[fill=green, line width=0.7086614168764338] (6.508333, -5.183333) rectangle (6.633333, -5.850000);
  \path[fill=green, line width=0.7086614168764338] (6.675000, -5.433333) rectangle (6.800000, -5.850000);
  \path[fill=green, line width=0.7086614168764338] (7.008333, -5.675000) rectangle (7.133333, -5.841667);
  \path[fill=green, line width=0.7086614168764338] (6.841667, -5.591667) rectangle (6.966667, -5.841667);
  \path[fill=green, line width=0.7086614168764338] (7.175000, -5.758333) rectangle (7.300000, -5.841667);
  \path[fill=green, line width=0.7086614168764338] (7.508333, -5.758333) rectangle (7.633333, -5.841667);
  \path[fill=green, line width=0.7086614168764338] (7.341667, -5.758333) rectangle (7.466667, -5.841667);
  \path[fill=green, line width=0.7086614168764338] (7.341667, -5.758333) rectangle (7.466667, -5.841667);
  \path[fill=green, line width=0.7086614168764338] (7.675000, -5.758333) rectangle (7.800000, -5.841667);
  \draw[draw=black, line width=0.7086614168764338] (3.300000, -0.641667) -- (3.300000, -5.858333);
  \draw[draw=black, line width=0.7086614168764338] (0.458333, -0.833333) -- (0.541667, -0.833333);
  \draw[draw=black, line width=0.7086614168764338] (0.458333, -1.666667) -- (0.541667, -1.666667);
  \draw[draw=black, line width=0.7086614168764338] (0.458333, -2.500000) -- (0.541667, -2.500000);
  \draw[draw=black, line width=0.7086614168764338] (0.458333, -3.333333) -- (0.541667, -3.333333);
  \draw[draw=black, line width=0.7086614168764338] (0.458333, -4.166667) -- (0.541667, -4.166667);
  \draw[draw=black, line width=0.7086614168764338] (0.458333, -5.000000) -- (0.541667, -5.000000);
  \draw[draw=black, line width=0.7086614168764338] (3.208333, -5.016667) -- (3.291667, -5.016667);
  \draw[draw=black, line width=0.7086614168764338] (3.208333, -2.516667) -- (3.291667, -2.516667);
  \draw[draw=black, line width=0.7086614168764338] (3.208333, -5.016667) -- (3.291667, -5.016667);
  \draw[draw=black, line width=0.7086614168764338] (3.208333, -0.850000) -- (3.291667, -0.850000);
  \draw[draw=black, line width=0.7086614168764338] (3.208333, -4.183333) -- (3.291667, -4.183333);
  \draw[draw=black, line width=0.7086614168764338] (3.208333, -1.683333) -- (3.291667, -1.683333);
  \draw[draw=black, line width=0.7086614168764338] (3.208333, -3.350000) -- (3.291667, -3.350000);
  \draw[draw=black, line width=0.7086614168764338] (6.050000, -5.000000) -- (6.133333, -5.000000);
  \draw[draw=black, line width=0.7086614168764338] (6.050000, -2.500000) -- (6.133333, -2.500000);
  \draw[draw=black, line width=0.7086614168764338] (6.050000, -5.000000) -- (6.133333, -5.000000);
  \draw[draw=black, line width=0.7086614168764338] (6.050000, -0.833333) -- (6.133333, -0.833333);
  \draw[draw=black, line width=0.7086614168764338] (6.050000, -4.166667) -- (6.133333, -4.166667);
  \draw[draw=black, line width=0.7086614168764338] (6.050000, -1.666667) -- (6.133333, -1.666667);
  \draw[draw=black, line width=0.7086614168764338] (6.050000, -3.333333) -- (6.133333, -3.333333);
  \draw[draw=black, line width=0.7086614168764338] (3.208333, -0.366667) -- (3.291667, -0.366667);
\end{tikzpicture}
  \end{center}
\caption{Frequency of occurrence of a name per id. The $i^{th}$ bar indicates how many times the name occurs with its $i^{th}$ id.}\label{plastic}
\end{figure}

\paragraph{Phase~2}
The previous phase has just extracted all possible name candidates for all id occurrences. Hence, $R1$ is very noisy and contains a large number of wrong candidates. For instance, some name candidates are not entity names, but rather descriptive elements, such as ``Price'', ``see also'', or ``plastic''.
This problem could be reduced by using a better entity tagger $f_t^\name$, but will ultimately always  appear.

Since our corpus of Web pages is large, we expect such non-specific names to appear uniformly over
several ids, whereas the correct names will accumulate on one id. As an example, Figure~\ref{plastic} shows three real distributions of names across ids from our experiments (Section~\ref{sec:experiments}). The chemical name ``amphetamine'' appears 120 times with 10 different ids. But it appears 99 times for one of these ids (which is the correct id of amphetamine). The non-specific names ``plastics'' and ``propionate'' appear more uniformly with different ids.

Our goal in Phase~2  is to identify names such as ``amphetamine'' that show a clear preference for one id. Formally, we count for each name~$n$ how often it appears with the id~$id$ in $R1$:
\[\fr_n(id) \defeq \card{\{ \langle id', n', s', u' \rangle \in R1 \mid n' = n,~id' = id\}}\]
\noindent Now, we attempt to identify names $n$ whose distribution $\fr_n$ shows a clear outlier.
We experimented with several outlier detection methods, which we detail in
Appendix~\ref{sec:outliers}. In the end, the following
technique works best. Let $id^1_n$ and $id^2_n$ be the ids with the highest and second highest value
for $\fr_n(\cdot)$, breaking ties arbitrarily.
The distribution $\fr_n(\cdot)$ is said to have an outlier
if $id^1_n$ appears in more than 30\% of the cases, and at
least 3 times more often than $id^2_n$.
\begin{align*}
  \fr_n(id^1_n) > & 0.3 \times \sum_i \fr_n(i)\\
  \fr_n(id^1_n)> & 3 \times \fr_n(id^2_n)
\end{align*}
This technique is robust enough to work across the board for all id types that
we considered. If a name appears with only one id, it is always considered an outlier.
Now, Phase~2 removes all names that show no clear outlier:
\[R2 \defeq \{\langle id^1_n, n, s, u \rangle \mid \langle id^1_n, n, s, u \rangle \in
    R1,~n\text{ has outlier } id^1_n\}\]
The result is a table $R2$ containing names that are specific to one id.

\paragraph{Phase~3} While we have now filtered out the insufficiently specific
names, entities may still have several names, some of which may be wrong.
To remove less likely names, we pick, for each id, the name that appears the most often. If the most frequent candidate names have the same frequency, we take the one with the highest score (ties on the score are resolved arbitrarily):
\begin{align*}
  R3' \defeq &~ \{\langle id, n, s, u \rangle \mid \langle id, n, s, u \rangle \in R2,~\fr_n(id)=\max_{n'}~\fr_{n'}(id)\}\\[-.1cm]
R3\phantom{'} \defeq &~ \{\langle id, n, u \rangle \mid \langle id, n, s, u \rangle \in R3',~s \text{~maximal for this } id\}
\end{align*}
The result table $R3$ of Phase~3 contains, for every id, a single name,
and the URLs of all Web pages where this entity was found. This is our final
entity database.

\subsection{Coverage}

If we wanted to build a comprehensive database of all entity ids on the Web, we would have to parse all existing Web pages. However, in practice, we only have access to a subset of pages that was found through crawling. Hence, our dataset is necessarily incomplete. It may happen, for example, that we see the same entity id over and over again in our crawl, instead of seeing new ids that we could add to our collection. In the worst case, we could crawl half of the Web, but see only a small fraction of the distinct entities that exist.

Fortunately, this is unlikely to happen if we assume the crawl is sufficiently random. To show this, we focus on the set $\calW$ of pages on the entire Web that mention at least one entity of type~$t$, and we consider $\calP' \defeq \calP \cap \calW$ the set of the input pages $\calP$ that mention some entity of type~$t$. We write $\alpha$ for
$\card{\calP'}/\card{\calW}$, and we assume that $\calP'$ is a subset of $\alpha \card{\calW}$ pages drawn uniformly at random in $\calW$.

We call $\calE$ the set of all different entities of type~$t$ appearing in $\calW$, and
$\calE' \subseteq \calE$ those appearing in $\calP'$. Intuitively, $\calE'$ are
the entities of $\calE$ that we can extract from our sample. We assume that some entity of $\calE$
occurs in strictly more than one page of $\calP$, and we claim:

\begin{theorem}
  For any fixed $0 < \alpha < 1$, the expected value of~$\card{\calE'}$ over draws of $\calP'$ is
  strictly greater than $\alpha \card{\calE}$.
\end{theorem}

In other words, if we crawl a random subset of $50\%$ of all Web pages
that mention entities, then we can expect to see more than $50\%$ of all
entities in our sample.

Intuitively, we show that, for any entity $e \in \calE$, the
probability of obtaining~$e$ is $\geq \alpha$, which can be seen by choosing
one arbitrary page~$p$ where $e$ occurs and noticing that the probability of
drawing~$p$ is at least $\alpha$. We now give the detailed proof:

\begin{proof}
  Call $x$ the expected value of $\card{\calE'}$.
  For each entity $e \in \calE$, define a random variable $E_e$ which is $1$
  if $e$ occurs in $\calP'$, and $0$ otherwise. It follows that $\card{\calE'} =
  \sum_{e \in \calE} E_e$, as the number of entities occurring in $\calP'$ is
  the sum, for each entity, of $0$ if it occurs in $\calP'$ and $1$ otherwise.
  Now, by linearity of expectation, $x$ is
  the sum of the expected values of the $E_e$, and, as $E_e \in \{0, 1\}$,
  this means that $x = \sum_{e \in \calE} q_e$, where $q_e$ is the probability that
  $E_e = 1$.

  We now show that $q_e \geq \alpha$ for all $e \in \calE$. Indeed, let us fix
  $e \in \calE$, and let $p_e \in \calW$ be an arbitrary page where $e$ occurs. Let $E_e'$
  be a random variable which is $1$ if $p_e$ occurs in $\calP'$, and $0$
  otherwise, and $q_e'$ be the probability that $E_e' = 1$. By definition $E_e'
  = 1$ implies that $E_e = 1$, so that $q_e \geq q_e'$. Now, the
  variable $E_e'$ can be modeled as a hypergeometric distribution with the
  following parameters: the number of draws is $\alpha \card{\calW}$
  (corresponding to $\card{\calP'}$), the number of successes is $1$
  (corresponding to the page $p_e$), and the population size is $\card{\calW}$.
  Indeed, $E_e'$ is the number of successes (here, $0$ if $p_e$ is not drawn and
  $1$ if it is drawn) when drawing $\card{\calP'}$ pages, without replacement,
  from the $\card{\calW}$ possible pages. Hence, from the probability mass
  function of the hypergeometric distribution, we deduce that $q_e' =
  {\card{\calW} - 1 \choose \card{\calP'} - 1} / {\card{\calW} \choose
  \card{\calP'}}$, which simplifies to $\frac{\card{\calP'}}{\card{\calW}}$,
  that is, $q_e' = \alpha$, so that $q_e \geq \alpha$.

  What is more, letting $e_0 \in \calE$ be an entity that occurs in strictly more
  than one page of $\calP$, we have $q_{e_0} > q_{e_0}'$. Indeed, if
  $p_{e_0}'$ is a
  page different from $p_{e_0}$ where $e_0$ occurs, there is a draw of $\calP'$ where
  we have $E_{e_0}' = 0$ but $E_{e_0} = 1$, namely, any draw where $p_{e_0}' \in \calP'$ but
  $p_{e_0} \notin \calP'$. Hence, as these additional draws have non-zero
  probability, $q_{e_0} > q_{e_0}'$.

  Now, putting it together, we have that $x = \sum_{e \in \calE} q_e > \sum_{e
  \in \calE} q_e'$, as the bound is strict for the term with $e_0$. By the above,
  this simplifies to $x > \alpha \card{\calE}$.
\end{proof}

\section{Implementation}
\label{sec:parsepage}

We now discuss the detailed implementation of Phase~1 of our method from Section~\ref{sec:approach}, where we parse a Web page to
extract records, ids, and name candidates.

\subsection{Parsing Web Pages}

\paragraph{Requirements} The Web pages that we consider are written in HTML, which can in theory be parsed to a DOM tree that represents the structure of the page. However, while HTML defines tags with nesting syntax and semantics, neither of them is always respected by Web site creators. In fact, a large number of HTML documents on the Web cannot be properly parsed into a DOM tree because they are not well-formed.

In addition, the structure of the DOM tree does not correspond to the page structure as seen by a human. For example, if a page contains several \textsc{h1} tags, then a human sees several sections. The DOM tree, however, contains just one parent node with \textsc{h1}-nodes and text-nodes in alternation as children. If the page discusses different entities, then it is likely that each entity falls within one \textsc{h1}-dominated block (see Section~\ref{scoremodels} for experiments on this).
The DOM tree, however, does not make this directly apparent.
More generally, many websites use HTML markup in a non-semantic way which leaves
little information in the DOM tree about the relationship between the elements
in the page.

This problem is well-known, and it is addressed by work on Web page segmentation. Most approaches, however, render the page visually~\cite{103,96,42,vips}, including layout or style sheet analyses. This is too expensive in our scenario, where we need a rough and rapid segmentation of pages at Web
scale. Other approaches use machine learning~\cite{44,41}, but there is no
training data in our scenario. Another method, MDR~\cite{41}, can detect tabular structures in Web pages. Our method is inspired by MDR, but more generally applies to Web pages without tabular structures. Most importantly, it is robust enough to work on Web pages without a proper DOM tree, and simple enough to run at Web scale on arbitrary input.

\paragraph{Frame trees} We segment the HTML page $p$ into a \emph{frame tree} (the name of which is not related to frames in HTML documents).
A frame tree looks like a DOM tree, but contains additional nodes for blocks that are introduced by \emph{separators}. A separator is a tag that starts a new paragraph, such as \textsc{h1,...,h6}, \textsc{hr},  \textsc{br}, sequences of \textsc{br}, and \textsc{p}. Any opening HTML tag starts a new frame, and any matching closing tag closes the current frame. If there is no closing tag in a scope of a parent frame, then the current frame is closed when the parent frame is closed. Additionally, every separator starts a new frame that ends at the next occurrence of a separator of equal or higher weight, or at the end of a parent frame.
\begin{algorithm}
\begin{framed}
   \textbf{function} parse(HTML document $d$, parent tag $t$):\\
   \emph{// $d$ is passed by reference so recursive calls may modify it}\\
   FrameTree result $\gets$ $\langle$tag: $t$, content: $[~]\rangle$\\
   \textbf{if}  $t$ is self-closing \textbf{then}  return(result)\\
   \textbf{while} {$|d|>0$}\\
   \-\ \-\ \-\ \-\ \-\ \-\ \-\ \-\ \textbf{if} {$d$ starts with tag $t'$}\\
   \-\ \-\ \-\ \-\ \-\ \-\ \-\ \-\ \-\ \-\ \-\ \-\ \-\ \-\ \-\ \-\ \textbf{if} {$t'$ is closing}\\
   \-\ \-\ \-\ \-\ \-\ \-\ \-\ \-\ \-\ \-\ \-\ \-\ \-\ \-\ \-\ \-\ \-\ \-\ \-\ \-\ \-\ \-\ \-\ \-\ \textbf{if} {$t'$ closes $t$}\\
   \-\ \-\ \-\ \-\ \-\ \-\ \-\ \-\ \-\ \-\ \-\ \-\ \-\ \-\ \-\ \-\ \-\ \-\ \-\ \-\ \-\ \-\ \-\ \-\ \-\ \-\ \-\ \-\ \-\ \-\ \-\ \-\ remove $t'$ from $d$\\
   \-\ \-\ \-\ \-\ \-\ \-\ \-\ \-\ \-\ \-\ \-\ \-\ \-\ \-\ \-\ \-\ \-\ \-\ \-\ \-\ \-\ \-\ \-\ \-\ \-\ \-\ \-\ \-\ \-\ \-\ \-\ \-\ \textbf{break}\\
   \-\ \-\ \-\ \-\ \-\ \-\ \-\ \-\ \-\ \-\ \-\ \-\ \-\ \-\ \-\ \-\ \-\ \-\ \-\ \-\ \-\ \-\ \-\ \-\ \textbf{end if}\\
   \-\ \-\ \-\ \-\ \-\ \-\ \-\ \-\ \-\ \-\ \-\ \-\ \-\ \-\ \-\ \-\ \-\ \-\ \-\ \-\ \-\ \-\ \-\ \-\ \textbf{if} {$t'\not\succ t$}\\
   \-\ \-\ \-\ \-\ \-\ \-\ \-\ \-\ \-\ \-\ \-\ \-\ \-\ \-\ \-\ \-\ \-\ \-\ \-\ \-\ \-\ \-\ \-\ \-\ \-\ \-\ \-\ \-\ \-\ \-\ \-\ \-\ remove $t'$ from $d$\\
   \-\ \-\ \-\ \-\ \-\ \-\ \-\ \-\ \-\ \-\ \-\ \-\ \-\ \-\ \-\ \-\ \-\ \-\ \-\ \-\ \-\ \-\ \-\ \-\ \-\ \-\ \-\ \-\ \-\ \-\ \-\ \-\ \textbf{continue}\\
   \-\ \-\ \-\ \-\ \-\ \-\ \-\ \-\ \-\ \-\ \-\ \-\ \-\ \-\ \-\ \-\ \-\ \-\ \-\ \-\ \-\ \-\ \-\ \-\ \textbf{end if}\\
   \-\ \-\ \-\ \-\ \-\ \-\ \-\ \-\ \-\ \-\ \-\ \-\ \-\ \-\ \-\ \-\ \-\ \-\ \-\ \-\ \-\ \-\ \-\ \-\ \textbf{break}\\
   \-\ \-\ \-\ \-\ \-\ \-\ \-\ \-\ \-\ \-\ \-\ \-\ \-\ \-\ \-\ \-\ \textbf{end if}\\
   \-\ \-\ \-\ \-\ \-\ \-\ \-\ \-\ \-\ \-\ \-\ \-\ \-\ \-\ \-\ \-\ \textbf{if}  $t\not\succ t'$  \textbf{then} \textbf{break}\\
   \-\ \-\ \-\ \-\ \-\ \-\ \-\ \-\ \-\ \-\ \-\ \-\ \-\ \-\ \-\ \-\ remove $t'$ from $d$\\
   \-\ \-\ \-\ \-\ \-\ \-\ \-\ \-\ \-\ \-\ \-\ \-\ \-\ \-\ \-\ \-\ \textbf{if} {$t'$ is separator}\\
   \-\ \-\ \-\ \-\ \-\ \-\ \-\ \-\ \-\ \-\ \-\ \-\ \-\ \-\ \-\ \-\ \-\ \-\ \-\ \-\ \-\ \-\ \-\ \-\ $f \gets$ [ parse($d$, $t'$) ]\\
   \-\ \-\ \-\ \-\ \-\ \-\ \-\ \-\ \-\ \-\ \-\ \-\ \-\ \-\ \-\ \-\ \-\ \-\ \-\ \-\ \-\ \-\ \-\ \-\ $f$.appendAll(parse($d$, $t'^*$).content)\\
   \-\ \-\ \-\ \-\ \-\ \-\ \-\ \-\ \-\ \-\ \-\ \-\ \-\ \-\ \-\ \-\ \-\ \-\ \-\ \-\ \-\ \-\ \-\ \-\ result.content.append($\langle$tag: $t'^*$, content: $f\rangle$)\\
   \-\ \-\ \-\ \-\ \-\ \-\ \-\ \-\ \-\ \-\ \-\ \-\ \-\ \-\ \-\ \-\ \textbf{else}\\
   \-\ \-\ \-\ \-\ \-\ \-\ \-\ \-\ \-\ \-\ \-\ \-\ \-\ \-\ \-\ \-\ \-\ \-\ \-\ \-\ \-\ \-\ \-\ \-\ result.content.append(parse($d$, $t'$))\\
   \-\ \-\ \-\ \-\ \-\ \-\ \-\ \-\ \-\ \-\ \-\ \-\ \-\ \-\ \-\ \-\ \textbf{end if}\\
   \-\ \-\ \-\ \-\ \-\ \-\ \-\ \-\ \textbf{else}\\
   \-\ \-\ \-\ \-\ \-\ \-\ \-\ \-\ \-\ \-\ \-\ \-\ \-\ \-\ \-\ \-\ read and remove text $s$ from $d$\\
   \-\ \-\ \-\ \-\ \-\ \-\ \-\ \-\ \-\ \-\ \-\ \-\ \-\ \-\ \-\ \-\ \textbf{if} $s$ is not whitespace \textbf{then} result.content.append($s$)\\
   \-\ \-\ \-\ \-\ \-\ \-\ \-\ \-\ \textbf{end if}\\
   \textbf{end while}\\
   \textbf{return} result
\end{framed}   
\caption{Building a frame tree}
\label{frames}
\end{algorithm}

Algorithm~\ref{frames} parses an HTML document recursively into
a frame tree. It is called initially with a dummy parent tag \textsc{doc}, and relies on a containment relation $\succ$ on tags, so that $t
\succ t'$ if a tag $t$ can contain a tag $t'$. For example,
\textsc{doc}$\succ$\textsc{html}$\succ$\textsc{body}$\succ$\textsc{div}. This order can be derived
from the HTML grammar. In addition, we introduce an artificial tag $t^*$ for
every separator tag $t$. For example, we introduce the tag \textsc{h1$^*$},
which will be the label for a frame that consists of a \textsc{h1}-header and
the following text. We extend $\succ$ to cover also these tags. For example, a
\textsc{h1}-frame can contain \textsc{h2}-frames:
\textsc{h1$^*$}$\succ$\textsc{h2$^*$}.
The algorithm yields a tree of frames, whose leaf nodes are text nodes. We call these nodes
\emph{text frames}. Algorithm~\ref{frames} will produce frame trees even if the page is not fully standards-compliant.

\paragraph{Example} Consider the following HTML document:

\begin{framed}\ttfamily\noindent<body>\\
$~~~~$<h1>Samsung Galaxy S4</h1>\\
$~~~~$$~~~~$<p>Id: <b>8806085725072\\
$~~~~$<h1>Accessories\\
$~~~~$$~~~~$<h2>Galaxy S4 Charging Cable</h2>\\
$~~~~$$~~~~$4047443213525\\
</body>
\end{framed}

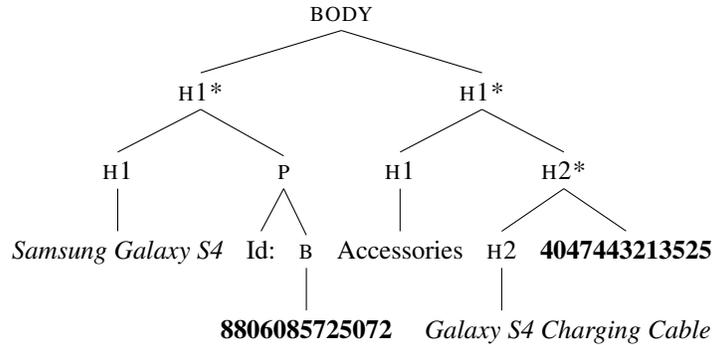
\begin{figure}
  \begin{center}
  \begin{tikzpicture}
    \Tree
      [.\html{body}
        [.\html{h1*}
          [.\html{h1}
            [.{\emph{Samsung Galaxy S4}} ]
          ]
          [.\html{p}
            [.Id: ]
            [.\html{b}
              [.{\is{\textbf{8806085725072}}} ]
            ]
          ]
        ]
        [.\html{h1*}
          [.\html{h1}
            [.Accessories ]
          ]
          [.\html{h2*}
            [.\html{h2}
              [.\is{~~~~~~~~~~~~~~~~~~~~\emph{Galaxy S4 Charging Cable}} ]
            ]
            [.{\textbf{4047443213525}} ]
          ]
        ]
      ];
  \end{tikzpicture}
  \end{center}
  \caption{Frame tree produced by Algorithm~\ref{frames}}
  \hspace{.25cm}
  \label{fig:frametree}
\end{figure}

This document is not correct HTML: there is no \texttt{<html>} or
\texttt{<head>} element, closing tags are missing, etc. Yet,
Algorithm~\ref{frames} is able to parse this document, yielding the frame tree
of Figure~\ref{fig:frametree} (omitting the dummy \textsc{doc} node). The figure shows ids in bold and the names that should be extracted in italics.

\subsection{Extracting Records}

Algorithm~\ref{frames} has produced a frame tree for an input Web page. We now show how to find \emph{records} in this tree. A record $r$ of type $t$ in a page $p \in \mathcal{P}$ is a subtree rooted at some node of the frame tree of $p$ that contains a single id of type $t$.
For this, a text frame must correspond exactly to one id, with no surrounding text.
We say that $r$ is a \emph{detail record} if it is the only record of $p$; otherwise, we call it a \emph{free record}. Intuitively, detail records occur when the entire page is concerned about the entity, and free records typically belong to listings of entities, such as lists or tables, or possibly free-floating descriptions of entities.

We find records on a page as follows. We apply the id validator $f_t^\id$ to all text frames of the frame tree, and mark the frames that are accepted by the validator and are therefore valid ids. If the page contains exactly one id, we have a detail record, which is just the root frame of the entire page.
If the page contains several ids, we traverse the frame tree in a depth-first search. As soon as we find a subframe that contains exactly one id, we construct a free record from that subframe. In our example in Figure~\ref{fig:frametree}, we would find two free records, rooted at each \html{h1*} node. Note that multiple occurrences of the same id in a page will always be allocated to different records.

Once all records in a page have been identified, the function $f_t^\name$ is applied to all text subframes of each record. This yields a set of candidate names per record. In our example in Figure~\ref{fig:frametree}, we would extract the candidate \emph{Samsung Galaxy S4} for the first id, and the candidates \emph{Accessories} and \emph{Galaxy S4 Charging Cable} for the second id.
The ids together with their candidate names make up the table $R1$ of our Phase~1.

\subsection{Discussion}

We remark that our approach deals correctly with many common structures in HTML documents that refer to entities. If, for example, each row of an HTML table contains an id, then each row will become a record in Algorithm~\ref{frames}. Similarly, for HTML lists, if each item in an HTML list contains an id, then each item will become a record, and so on for any other repetitive structure, no matter which tag is used to delimit the items (\textsc{table}, \textsc{ul}, or any other tag). Conversely, a \textsc{table} tag that does not actually describe a table will not confuse the algorithm. This makes the algorithm robust enough to run on arbitrary Web sites.

If each row of a table corresponds to an entity, one could expect that entity names will all be in the same column. Conversely, a column that always contains the same word is unlikely to contain an entity name. As our approach is generic, Phase~1 extracts all candidate names agnostically. Then, Phase~2 (see Section~\ref{sec:approach}) will remove globally frequent names. Thus, Phase~2 will have the same effect as discarding frequent words from table columns, just that it operates on a global table. Finally, Phase~3 will choose the most frequent name for an id.
Thus, Phase 2 and Phase 3 together act like the TF-IDF mechanism in information retrieval.

\section{Experiments}
\label{sec:experiments}

\begin{table*}[t]
\caption{Total number of items, accuracy, and recall after each phase}
\label{accuracy-name}
  \begin{tabularx}{\linewidth}{X@{\hskip 20pt}ccc@{\hskip 32pt}ccc}
\toprule
    & \multicolumn{3}{c}{\bf GTIN items}& \multicolumn{3}{c}{\bf CAS items} \\
        & {\bf Num.} & {\bf Acc.} & {\bf Rec.} & {\bf Num.} & {\bf Acc.} & {\bf Rec.} \\
	\midrule
Phase~1 & 3,929,312 & 38\%$\pm$8\% & 60\% & 241,602 & 76\%$\pm$6\% & 80\%  \\
Phase~2 & 2,550,703 & 76\%$\pm$8\% & 48\% & 235,779 & 86\%$\pm$5\% & 76\%  \\
Phase~3 & 2,550,703 & 82\%$\pm$7\% & 50\% & 235,779 & 96\%$\pm$3\% & 78\%  \\
\bottomrule
\end{tabularx}

\smallskip

  \begin{tabularx}{\linewidth}{X@{\hskip 20pt}ccc@{\hskip 20pt}ccc}
\toprule
    & \multicolumn{3}{c}{\bf DOI items}& \multicolumn{3}{c}{\bf Email items} \\
        & {\bf Num.} & {\bf Acc.} & {\bf Rec.} & {\bf Num.} & {\bf Acc.} & {\bf Rec.} \\
	\midrule
Phase~1 & 1,167,810 & 52\%$\pm$8\% & 50\% & 13,625,860  & 90\%$\pm$6\%  & 63\%\\
Phase~2 & 1,038,950 & 68\%$\pm$9\% & 45\% & 13,625,860 & 90\%$\pm$6\% & 63\% \\
Phase~3 & 1,038,950 & 73\%$\pm$8\% & 47\% & 13,625,859 & 90\%$\pm$6\% & 63\% \\
\bottomrule
\end{tabularx}

\end{table*}

This section presents our experimental results. We first describe our setup.
Our input is the ClueWeb corpus, a large Web crawl. We target the English
portions of ClueWeb09 and ClueWeb12. In total, our corpus is around 35 TB in
size, and contains 1.2 billion Web pages.
We ran our approach on this corpus
for the id types GTIN, CAS, and DOI, as well as the pseudo-id \emph{email}\footnote{Since many people share the same name, we skipped Phase 2 for email addresses.}.
We used the simple NER modules described in Appendix~\ref{app}.

We implemented our algorithm in a Map-Reduce framework. Phase~1 is highly parallelized, with every mapper extracting from a different part of the corpus. Phase~2 and Phase~3 are classical grouping tasks, which come natively with the Map-Reduce framework.
The extraction process took 10 hours on a Hadoop cluster of 10 nodes, with the total capacity of 80 map-reduce tasks, amounting to an average of about 3,000 pages, or 100 MB, processed per second and per node.

\subsection{Entity Extraction}

\paragraph{Quality} To verify the quality of our name extraction, we produced a gold standard set of ids and entity names. For this purpose, we randomly sampled, for each type, 200 occurrences of ids in pages from our Web corpus. We annotated each id manually with its correct name in the page. Then, we compared the output of each phase to this gold standard.
We measured accuracy and recall and the total number of items after each phase.
For the first two phases, as there are multiple name candidates per id, we pick one name at random for each id, to simulate a guess.
Table~\ref{accuracy-name} shows our results.
To perform the evaluation, we considered each id, and compared the assigned name to the correct name from the gold standard. Accuracy is the proportion of correct names. Recall is the proportion of correct names that was correctly assigned.
To make sure that our results are statistically significant, we compute the Wilson score interval \cite{wilson} for each evaluation\footnote{This score allows estimating the true ratio of correct names in a population of arbitrary size from a sample, if the sample is drawn randomly (as in our case).}.

As we can see, Phase~1 cannot find
all entities that have an id. It also has a low accuracy,
because it produces many name candidates, one of which is chosen at random for the evaluation.
This is essentially what a naive \textbf{baseline} algorithm would do: it would extract all candidate  names on a Web page, and assign one name at random to each id on that page. As we see, the performance is mediocre, with a accuracy of 38\% for GTINs.
The beauty of our approach is that even with such mediocre results in the first phase, the second phase filters out erroneous names, increasing the accuracy and yielding very good results overall. The third phase guesses the correct name for each entity, which increases the accuracy even further -- up to 96\%. The systematic cleaning of Phase~2 and Phase~3 can double the accuracy for products.
The recall varies through the phases, since the set of candidate names per id shrinks, which may increase or decrease the recall. In the end, our method assigns the correct name for 83\% of the products, 96\% of the chemical substances, 73\% of the documents, and 90\% of the email addresses.

\begin{table*}[t]
\caption{Richest sources for entities of various entity types\label{tab:domains}}
\begin{center}
\begin{tabular}{p{6cm}r}
\toprule
{\bf Product sources} & {\bf Items}\\
\midrule
\url{www2.loot.co.za}		& 304,431 \\
\url{www.books-by-isbn.com}& 50,683\\
\url{gtin13.com} & 26,834\\
\url{en.wikipedia.org}& 21,873\\
\url{www.buchhandel.de}& 18,264\\
\bottomrule
\end{tabular}

\smallskip

\begin{tabular}{p{6cm}r}
\toprule
{\bf Chemical sources} & {\bf Items}\\
\midrule
\url{www.chembuyersguide.com} &129,211\\
\url{www.chemnet.com} & 22,061\\
\url{www.lookchem.com} &	12,354\\
\url{www.seekchemicals.com} & 7,326	\\
\url{www.tradingchem.com}	&4,769\\
\bottomrule
\end{tabular}

\smallskip

\begin{tabular}{p{6cm}r}
\toprule
{\bf Document sources} & {\bf Items}\\
\midrule
\url{wwwtest.soils.org} &	\phantom{1}20,635 \\
\url{www.plosone.org}	& 19,261	\\
\url{www.citeulike.org}	 & 13,491 \\
\url{www.astm.org} & 10,020 \\
\url{bja.oxfordjournals.org} & 9,030 \\
\bottomrule
\end{tabular}
\end{center}
\end{table*}

\paragraph{Quantity} As Table~\ref{accuracy-name} shows, our database contains 2,550,703 products, 235,779 chemical substances, 1,038,950 documents, and 13,625,859 email addresses. 55.6\% of our products are books. The other products include things as diverse as office supply items, DVDs, or USB cables. While email addresses are pseudo-ids, and we cannot guarantee uniqueness in this case, all other ids are unique. This means that our numbers provide a lower bound for the number of chemical substances, unique documents, books, and commercial products on the Web.
Our database is also orders of magnitude larger than other public databases.

To estimate the coverage of our database, we compared our data to the YAGO KB
\cite{yago}, focusing on books, whose ids (ISBN codes) are known to YAGO. YAGO
contains 11,271 books with an ISBN. Of these, 1,662 appear in our database.
We assume that we missed some of the YAGO books because Wikipedia, the source of YAGO, is not necessarily entirely in our corpus; furthermore, our cleaning phases may remove correct book candidates.
By contrast, our database contains 1.4 million books that are unknown to YAGO.

\subsection{Extensions}\label{extension}

\paragraph{Attribute extraction} Entities can have certain attributes. Commercial products, e.g., can have a price, and chemicals have a molecular formula. To show that our approach could in principle be extended to extract also the attributes of an entity, we consider the extraction of molecular formulae for chemical substances (e.g., ``Cd5Cl(PO4)3'').  We build a regular expression that accepts any sequence of names and digits, where each name has to be a valid chemical element name from the periodic table. This yields an \emph{attribute-finder}, which works analogously to a name finder, and which runs through all 3 phases. With this approach, we could extract 1,662 chemical formulae. We picked a random sample of 50 values, and checked them manually. We obtain an accuracy of 93\%$\pm$6\%\footnote{The center of the Wilson interval, 93\% is the estimated ratio, which may differ from the percentage of correct evaluations.}. This analysis just serves to showcase that our approach could be extended to extract
also attributes of the entities.

\paragraph{Other id types} In order to estimate the potential of our approach for other types of ids, we implemented the name finders and id checkers for ICD-10 disease codes,  OMIM disease codes, VATIN company tax codes for France, and MESH chemical codes. On our corpus, the algorithm returned 2,661 distinct ICD-10 diseases, 2,418 distinct MESH chemicals, 7,521 OMIM diseases, and 240 VATIN companies. On the French part of ClueWeb~2009, in contrast, we found 2,233 distinct VATIN companies. These numbers make us believe that our approach can be extended to different types of ids and to different corpora.

\paragraph{Extrinsic application} Our data can be seen as a basic knowledge base (entities, labels, ids, types) and thus can be used in different scenarios employing KBs.
As a proof of concept we merged our data with the YAGO KB \cite{yago} and fed the data to the PATTY system~\cite{patty}.
PATTY finds typed textual patterns between named entities in a corpus, such as ``X was born in Y''. Run with YAGO on the New York Times corpus, it produces 99k such patterns. If our database is added, it produces 17k new patterns. Manual inspection shows that these come mostly from person names (``X called her husband~Y''), but also from products (``X to buy DET ADJ cellphone like DET Y'').

\section{Analyses}
\label{sec:analyses}

Our dataset is a huge resource of Web objects, which can give rise to different analyses~-- much in the spirit of Culturomics \cite{culturomics}. The following experiments illustrate this.

\begin{table}[t]
\caption{Common person names}
\centering
\label{commonNames}
  {\renewcommand{\tabcolsep}{3pt}
  \begin{tabular}{lr}
\toprule
\multicolumn{2}{l}{\bf Family names} \\
\midrule
Smith &  84,376 \\
Johnson &55,277 \\
Brown  & 46,499 \\
Jones  & 45,322 \\
Williams & 43,492\\
\bottomrule
\end{tabular} $~~~~$\begin{tabular}{lr}
\toprule
\multicolumn{2}{l}{\bf Given names} \\
\midrule
 John  &  238,446 \\
David  & 207,931  \\
Michael& 155,880  \\
 Mark    &117,755 \\
Robert & 109,814  \\
\bottomrule
\end{tabular} $~~~~$ \begin{tabular}{lr}
\toprule
\multicolumn{2}{l}{\bf Full names} \\
\midrule
John Smith & 1,969 \\
David Smith & 1,484\\
John Doe & 1,371   \\
Michael Smith & 990\\
David Brown & 899  \\
\bottomrule
\end{tabular}
}

\end{table}

\begin{table}[t]
    \begin{minipage}{.52\linewidth}\noindent
\caption{Top companies by \mbox{production}}
\label{brandTable}
{\renewcommand{\tabcolsep}{3pt}
  \begin{tabularx}{\linewidth}{Xlr}
\toprule
{\bf Company} & {\bf Prefix} & {\bf Products} \\
\midrule
 Bernat & 0057355 & 1,116 \\
 Panasonic & 0037988 & 929 \\
 Lion & 0023032 & 927 \\
 Nikon & 0018208 & 829 \\
\bottomrule
\end{tabularx}
}
\end{minipage}
\hfill
    \begin{minipage}{.44\linewidth}
      \caption{Top e-mail domains\protect\phantom{p}}
\label{mailDomains}
{\renewcommand{\tabcolsep}{3pt}
  \begin{tabularx}{\linewidth}{Xr}
\toprule
{\bf Domain} & {\bf Addresses}\\
\midrule
gmail.com & 304,236 \\
yahoo.com & 290,292 \\
hotmail.com & 281,498 \\
aol.com & 259,769 \\
\bottomrule
\end{tabularx}
}

\end{minipage}
\end{table}

\begin{table}[t]
\caption{Countries by production}
\label{countryNumProducts}
\centering
  \begin{tabular}{l@{\hskip 28pt}r}
\toprule
{\bf Country} & {\bf \#Products}\\
\midrule
USA &  1,024,219 \\
UK & 59,542      \\
Germany  & 26,949\\
Japan  & 17,353  \\
France & 12,845  \\
\bottomrule
\end{tabular} $~~~~$ \begin{tabular}{l@{\hskip 20pt}r}
\toprule
{\bf Country} & {\bf GDP (trillion)}\\
\midrule
 USA  &  14.99 US\$    \\
 China  & 7.20 US\$    \\
 Japan & 5.87 US\$     \\
 Germany    & 3.60 US\$\\
 France & 2.77 US\$    \\
\bottomrule
\end{tabular}
\end{table}

\subsection{Resources}

Our dataset can identify Internet domains that are particularly rich in different types of Web entities. In Table~\ref{tab:domains}, we show the best data sources that we could determine for email addresses, chemicals, and documents. This analysis could help steer information extraction approaches to target domains that are rich in the desired items. Most notably, \texttt{amazon.com} is not among the most common domains. We assume that, if it were added to the crawl, it would multiply the number of products that we would find.

We also computed the most frequent email providers occurring in the email addresses that we collected (Table~\ref{mailDomains}). Our email addresses come mostly from Gmail, followed by Yahoo!\ and Hotmail. These are indeed the top three email providers, as determined by the Techspot magazine~\cite{techspot}.

\subsection{People}

\paragraph{Common names} Our extraction found 13~million email addresses with an associated person name. However, email addresses are only \emph{pseudo-ids}: a person can have several email addresses, and so we may not conclude that we found 13 million people. However, we may assume that the number of email addresses that a person owns is independent of their name. Therefore, we can compute the most common given names and the most common family names on the Web (Table~\ref{commonNames}).
The popular first names that we found correspond roughly to the frequent English names. Our top 50 male given names cover 43 of the top 50 male given names of the US 1990 census data \cite{census}. We also mined the most common complete names, with
``John Smith'' being the single most common name.

\paragraph{Gender}
We extended our analysis to finding the gender of the people on the Web. By intersecting our set of given names with the US census data about common female and male first names, we can assign a gender to each person name. For the 331 unisex names, we attributed both genders proportionally, based on the name frequency statistics, to take advantage of any gender priors on them. We find that women are slightly under-represented: out of 11.6 million names in our database whose gender we could identify, only 47\% were female. 1,990,290 names were not recognized as American names.

\subsection{Commercial Products}

\paragraph{Company names} The first 4-7 digits of the GTIN identify the company that produced the product. Unfortunately, there is no publicly available database that maps these prefixes to company names. Therefore, we resorted to the following method. We assume that the product titles often contain the company name. We grouped the GTINs by their 4-digit prefix. Then, we computed the word that was most frequent within a group, but infrequent outside the group.
A manual analysis on a random sample of 80 products shows that this method can indeed identify the company name (or at least part of it) correctly in 83\% of the cases, with a Wilson score interval of $\pm$7\%. Table~\ref{brandTable} shows the companies with the most products.

\paragraph{Countries} The first 3 digits of a GTIN identify the country of production of a product. To conduct this analysis, we extracted the GTINs from the entire ClueWeb corpus (not just the English one). We calculated the number of unique items produced in different countries, and show the top countries in Table~\ref{countryNumProducts}.  If we compare the ranking to the list of countries by GDP (as provided by the World Bank~\cite{wb}), we find a remarkable overlap. Our top 5 countries cover 4 out of the 5 countries with highest GDP. To investigate this similarity further, we built a vector that contains, for each country, the number of products that we could find. We built another vector that contains, for each country, its GDP. We find a cosine similarity of 79\% between these vectors. We take this as an indication that the GTINs can serve as a proxy of the productivity of a country.

\begin{figure}[t]
\fbox{\includegraphics[width=.99\linewidth, clip=true, trim=0 10cm 3cm 10cm]{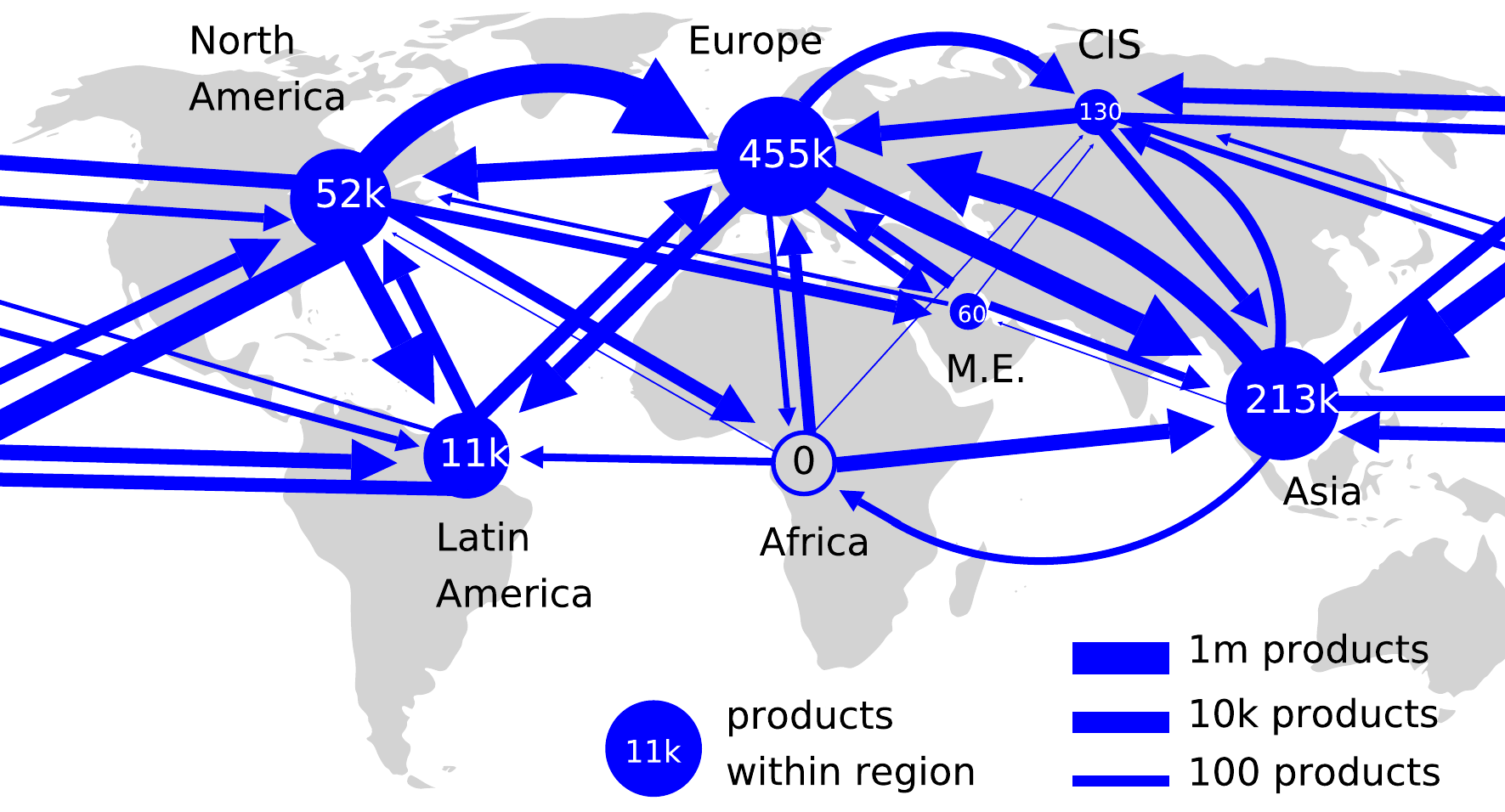}}
\caption{Intra- and inter-regional trade (log-scale)}
\label{map}
\end{figure}

\paragraph{Global trade} While the GTIN indicates where a product was produced,
the top level domain of a page where the product appears probably identifies a country where the product is sold. This allows us to trace which countries export to which other countries. We grouped the countries by the regions that the World Trade Organization (WTO) uses~\cite{wto}. We took again the GTINs from the entire ClueWeb, and plotted the trade flow on a map\footnote{\href{http://commons.wikimedia.org/wiki/File:World_map_without_Antarctica.svg}{by Wikicommons user \emph{E\_Pluribus\_Anthony}}} (Figure~\ref{map}).
``CIS'' stands for the Commonwealth of Independent States, which consists of the former Soviet Republics. The size of the circles corresponds to the number of products produced in and advertised within one region. The size of the arrows corresponds to the number of products produced in one region and advertised in
another one. The scale is logarithmic. We found no products produced in Africa and advertised in Africa. The other quantities correspond roughly to what one would expect: Europe, North America, and Asia are the dominant exporters.

We compared our analysis to the true flow of products between the regions, as estimated by the WTO~\cite{wto2}. For each region, we construct a vector that contains the number of exported products for each target region. We compare this vector to the vector of exported merchandise in US\$ from the WTO. We computed the cosine similarity of the vectors for each region, and found remarkable overlap. The similarity is lowest (49\%) for the Middle East. We hypothesize that this is due to the fact that the Middle East exports many commodities that are not sold by GTIN. For Europe, Asia, and the CIS, however, the values are over 95\%.

\section{Conclusion}
\label{sec:conclusion}

In this paper, we have shown how to harvest entities with unique ids systematically from the Web. By making use of the properties of ids, we could extract entity names with an accuracy of 73--96\%. This allowed us to create a database of 13 million email addresses with their name, 235 thousand chemical substances, 1 million documents,
1.4 million books, and 1.1 million other commercial products.
We believe that this dataset is the first public database that contains so many such items in a canonicalized manner. We have shown possible uses of this database by conducting a number of analyses, which include the frequent names of people, and the flow of trade in the world. We expect that many more exciting experiments can be conducted with our data.

We believe that our methodology is applicable not just to the types of entities that we picked here as examples, but also to a broad range of other entities. It should be possible to extract banks, companies, audiovisual material, or possibly even social network ids.
This will make the Web more and more semantic, and thus help making the Internet ever more useful.

 All data and analyses are publicly available at \url{http://resources.mpi-inf.mpg.de/d5/ibex}.

 \bigskip

\paragraph{Acknowledgements} We are grateful to Pierre Senellart for his useful feedback. This
work has been partly funded by the Télécom
ParisTech Research Chair on Big Data and Market Insights, as well as by the Labex
DigiCosme (project ANR-11-LABEX-0045-DIGICOSME) operated by the French ANR as part of the
program ``Investissement d'Avenir'' Idex Paris-Saclay (ANR-11-IDEX-0003-02).

\newpage

\appendix

\section{Scoring Name Candidates}
\label{scoremodels}

We now discuss several design alternatives for the scoring model used in Phase~1. We compare the performances of the design alternatives on test data. For this, we use the same experimental setup as in our final experiments (see Section~\ref{sec:experiments}).

\paragraph{Score design} In Phase 1, each pair of an entity id and an entity name receives a real-valued score. The score can also be NIL, meaning that the candidate name is completely removed from the list.
To evaluate different design alternatives for scoring models, we compared them against a gold standard constructed in the following way. We first ran just the id validator $f^\id_t$ for all id types $t$ on all text frames of all pages of our corpus. We sampled a set of 200 id occurrences randomly for each type (different from those of Section~\ref{sec:experiments}). We manually inspected the Web pages where these ids occurred, and extracted the correct entity name for each id by hand.

Now, to evaluate the performance of a scoring model against this gold standard, we run Phase~1 of our
extraction on all pages of $\calP$, and rank for every id all the candidate names by decreasing
score, breaking ties arbitrarily. (Note that the scoring model may also remove
a candidate entirely from the list.) Our goal is to design a scoring model that retains the correct name candidates, and ranks them closer to the top.
Formally, we compute the \emph{precision at rank $i$} as the proportion of ids with non-empty rankings where the correct name is among the first~$i$ ranked candidates. The \emph{recall at rank $i$} is
the proportion of ids where the correct name is among the first $i$. We tried
different scoring models. Table~\ref{scores} shows the results for GTINs,
choosing for each scoring model the rank that maximises the precision (the results for the other types are comparable).

\begin{table}[t]
\caption{Precision and recall of different scoring models for GTINs on different record types}\label{scores}
{
  \renewcommand{\tabcolsep}{5pt}
\begin{tabularx}{\linewidth}{lXrrr}
\toprule
{\bf Type} & {\bf Scoring model} & {\bf Prec.} & {\bf Rec.} & {\bf Rank} \\
\midrule
Detail & random                                      & 86\% & 25\% & 275\\
Detail & title                                           & 71\% & 16\% & 17\\
Detail & order                                         & 84\% & 24\% & 187\\
Detail & order + title                               & 70\% & 16\% & 6\\
Detail & order + distance                         & 84\% & 24\% & 78\\
Detail & order + distance + style4             & 68\% & 15\% & 3\\
Detail & order + distance + style4 + title   & 75\% & 10\% & 1\\
\midrule
Free & first3            & 80\% & 29\% & 9\\
Free & first3 + style & 85\% & 29\% & 9\\
\bottomrule
\end{tabularx}
}
\end{table}

\paragraph{Models for detail records} We now present the various scoring models that we
evaluate, looking only at detail records. Our first scoring model, \emph{random}, assigns the same value $1$ to all candidates. Thus, the precision and recall reflect what a random assignment of extracted entity names to ids would produce. Predictably, the performance is not very impressive, and the maximum precision of 86\% is achieved only at rank 285 (we do not achieve a precision of 100\% because the NER modules are not perfect).

Our next scoring model, \emph{title}, removes a name if the name does not occur in the \textsc{title} tag of the page (and assigns $1$ otherwise). This decreases recall slightly, but improves the rank with maximal precision drastically, to 17. We observed that the name of an entity occurs before the id in the large majority of cases. Hence, our next model, \emph{order}, removes a candidate if it appears after the id in the HTML file (and assigns $1$ otherwise). This decreases the recall only slightly, but improves the rank with maximal precision to 187.
The table shows
that a combination of the \emph{order} and \emph{title} feature improves the rank with maximal precision even to 6.

Through manual inspection, we found that name candidates that are closer to the id are more likely to be the correct name. Hence, our next scoring model, \emph{distance}, scores a name by the negative distance between the name and the id. If, e.g., there are 5 name candidates that lie between the current candidate and the id, then the score will be -5.  Combining \emph{order} with \emph{distance} (by using the distance score, and removing a candidate if \emph{order} says so) has a very positive effect on the rank with maximal precision.

We also observe that the tags that contain the name candidate play an important role. For example, a headline \textsc{h1} is most likely the entity name, even if it is far away from the id. Hence, our next scoring model assigns a score of 1 for ``hiding tags'' such as \textsc{small} or \textsc{strike}, a score of 2 for plain text, 3 for ``highlighting tags'' such
as \textsc{b} or \textsc{i}, and 4 for headers such as \textsc{h1}. Any finer scale of styles did not seem to have any additional effect. We found that we achieve the best results if we remove all candidates that do not have a tag score of 4. This is what the scoring model \emph{style4} does. By combining all of these scoring models, we achieve very good precision already at rank~1. This combined model is what we will use for detail records.

\paragraph{Models for free records} The \emph{title} feature cannot be applied to free records. However, a variant of \emph{order}, which restricts the extraction to the first 3 text frames of each record, yielded good precision (shown as \emph{first3}). The \emph{style} score, likewise, helps. If we combine these two scoring models (by using the full style score as described above, and by removing any candidates beyond the 3rd text frame), we obtain very good precision at a very small rank already. This is the scoring model that we use in the experiments for free records.

\section{Outlier Detection}
\label{sec:outliers}

We now discuss our choice of how to detect outliers in Phase~2. Our goal is to exclude names that are associated with many different ids, as we expect them to be general words rather than entity-specific names. Like in Appendix~\ref{scoremodels}, we rely on experiments to identify the best design choices.

\begin{figure}
  \centering
\includegraphics[width=\linewidth]{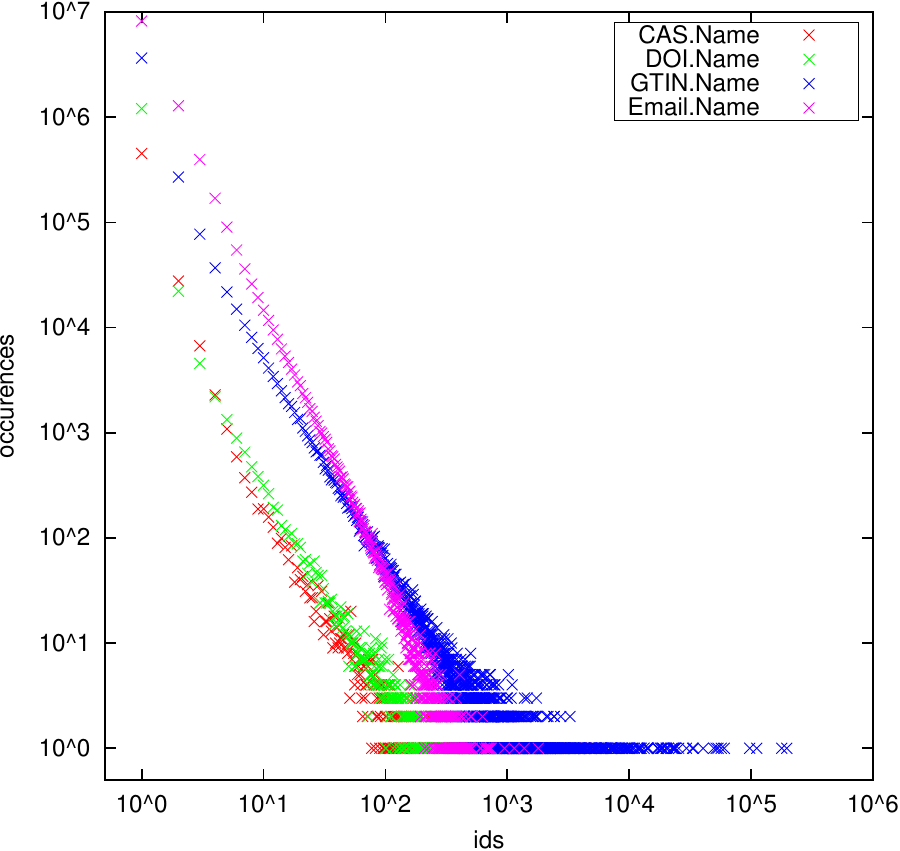}

\caption{Number of ids with which each particular name occurs in a log-log scale. A point $(x, y)$ means that there are $y$ names that occur with $x$ ids.}\label{loglog}
\end{figure}

We first ran Phase~1 on the entire corpus, and collected the ids with their name candidates. Figure~\ref{loglog} shows the number of ids that each particular name occurs with on a log-log scale.  Most names occur with only one id. However, there are names that occur with multiple ids, and not all names that occur with multiple ids are noise. For example, the names ``plastics'', ``amphetamine'', and ``propionate'' each occur with 10 ids. Figure~\ref{plastic} shows how often these names occur with each of their ids. Only ``amphetamine'' is a correct name for a chemical substance. It is clearly associated to one id, with the other occurrences being noise. ``plastics'' and ``propionate'' are more uniformly distributed across the ids, and are not names for chemical ids. Our
goal is to distinguish a correct name such as ``amphetamine'' from the incorrect names.

\paragraph{Schemes} There are many different methods to detect outliers. These include the Z-test, the Grubbs test, the Dixon test, and the Box-plot rule (see, e.g., \cite{outlier1, outlier2} for surveys). We looked into all four methods, but none of them is applicable in our setting. The reason is that we need to find an outlier already for a very small set of distinct ids (on average 11) without any knowledge of the underlying probability distribution.

Thus, we formalized our intuitive definition of an outlier. An outlier for
a name is the id that has the highest frequency, such that its frequency is
above a minimal threshold and there is no second id with comparable frequency.
This means that an id is an outlier if there is no second id, or if:
\[f_1> pn~~\wedge~~i \times f_2<f_1\]
Here $f_k$ is the frequency of the id with the rank $k$, $n$ is the total number
of id occurrences, and $i \in \mathbb{N}$ and $p \in [0,1]$ are parameters. We
now describe how we chose those parameters.

\paragraph{Parameter tuning}
To find these parameters, we used again the sample of id occurrences from
Appendix~\ref{scoremodels}. Each id in this sample has been mapped manually to
the correct name, giving us a set of \emph{correct} names. By contrast, we manually
selected 100 general names per id type from the output of Phase~1 (for example,
``Pet supplies'' from the list of GTIN name candidates), giving us a set of
\emph{incorrect} names. We considered all ids for which the sample names
occurred, giving us, for every correct and incorrect name, a distribution of
ids, as in Figure~\ref{plastic}. We evaluated which choices of~$i$ and~$p$ did
the best job at classifying these distributions on this test set.

We varied $i$ between 0 and 20, and $p$ between 0 and 0.3, and measured the
precision and recall when classifying good and bad names. In general, precision
increases with growing $i$ and $p$, and recall decreases. Our focus is on
precision, and hence we required the precision to be at least $95\%$. By varying
$i$ and $p$, we found that the combination $i=3$ and $p=30\%$ is a sweet-spot,
which achieves a precision of~$95\%$ and a recall of~$25\%$.

\section{NER Modules}\label{app}
We describe here one possible implementation of the NER modules $f^\id_t$ and
$f^\name_t$ from Section~\ref{sec:approach} for $t \in \{$GTIN, CAS, Email, DOI$\}$. These modules are the \emph{input} of our method, and any NER modules can be used in place of the ones described here.

\paragraph{GTINs} A GTIN contains 14 digits: one digit for the packaging level,
3 digits for the country, 4-7 digits for the company, and the remaining digits
for the product. The last digit is a check digit. $f^\id_\gtin$ checks the
length of the sequence, and validates the check digit. $f^\name_\gtin$ is a
validator that accepts the input string, if it starts with a letter or number,
contains a word of at least 4 characters, and contains no more than 250
characters. Books have GTINs that start with ``978''. To avoid that we take the
author of a book as its title, $f^\name_\gtin$ aggressively rejects candidates that look like author names or lists of author names. If the candidate string contains a given name or a single-letter abbreviation, or if more than one third of the tokens in the string are commas, the validator rejects it. We compiled a dictionary of first names from the 1990 US Census~\cite{census} and the Balie project \cite{balie}, from which we
removed some common words (such as ``China'').

\paragraph{CAS} A CAS number consists of 3 parts, separated by hyphens, where
the last part is a check digit. Hence, $f^\id_\cas$ checks the syntactic form
of the id, and validates the check digit. $f^\name_\cas$ is a validator that accepts the input string if it contains a word of at least 4 characters and is not more than 250 characters long. It rejects a candidate if it contains a chemical formula or any character other than alphanumeric characters, brackets, quotes and hyphens.

\paragraph{Email} $f^\id_\myemail$ checks whether the input consist of a local
part, followed by the @-sign and a domain name. $f^\name_\myemail$ has to recognize person names such as ``John Smith'', ``Smith, John'', and ``Dr. John Smith''. The literature has developed sophisticated approaches to this end \cite{sunita}. Here, we use a simple name finder that scans the input string and returns all substrings that follow the pattern ``\emph{first middle last}'' or ``\emph{last, first middle}''. Here, \emph{last} and \emph{middle} match any, possibly hyphenated, capitalized word. \emph{first} matches any combination of first names from our dictionary of first names.
Since we were only interested in personal email addresses and not in
organizations, Web administrators, or service providers, $f^\name_\myemail$  returns only person names that overlap with the email address. \ignore{To see whether a person name matches an email address, it removes all words of the name from the email address. If at most one letter remains before the '@', it accepts the name. The one letter is to account for middle names in email addresses.}

\paragraph{DOIs} $f^\id_\doi$ just verifies whether the id follows the pattern
of a numeric prefix followed by a slash and a sequence of characters. Document
titles are often not marked up separately, but occur in plain text. Therefore,
$f^\name_\doi$ has to search candidates for the document title in the input string.
\ignore{There is ample work on extracting the title of a Web page~\cite{106,91,93,92}. Here, however, we are not interested in extracting the title of a Web page, but in extracting the title of a document that is \emph{mentioned} in a Web page.
}
There is some work on this task (known as \emph{bibliographic reference parsing}, e.g. \cite{refparse}), but these approaches can only extract from homogeneously formatted lists of references. In our case, in contrast, the names appear as an arbitrary sub-sequence of plain text.
$f^\name_\doi$ splits the string by the separators \emph{.;"?!}. It accepts a substring as a candidate, if it contains at least 4 words. As with book titles, we exclude author names.

\bibliographystyle{abbrv}
\bibliography{main}
\end{document}